\documentclass[pre,showpacs,showkeys,preprintnumbers,amsmath,amssymb,superscriptaddress,twocolumn,nofootinbib]{revtex4}

\usepackage{graphicx}
\usepackage{comment} 
\usepackage[utf8]{inputenc}
\usepackage{xcolor}
\usepackage{color}
\usepackage{bm}
\usepackage{amsfonts}
\usepackage[colorlinks=true,linkcolor=blue,citecolor=red]{hyperref}%

\usepackage{url}
\usepackage{epstopdf}

\def\x{\bm{x}}

\def\l{\ell}

\begin{document}

\title{Mean first-passage time to a small absorbing target \\ in three-dimensional elongated domains}

\author{Denis S. Grebenkov}
\affiliation{Laboratoire de Physique de la Mati\`ere Condens\'ee (UMR 7643), \\ 
CNRS -- Ecole Polytechnique, IP Paris, 91120 Palaiseau, France}
\email{denis.grebenkov@polytechnique.edu}

\author{Alexei T. Skvortsov}
\affiliation{Maritime Division, Defence Science and Technology Group, 506 Lorimer Street, Fishermans Bend, Victoria 3207, Australia}
\email{alex.skvortsov@dst.defence.gov.au}

\date{\today}

\begin{abstract}
We derive an approximate formula for the mean first-passage time
(MFPT) to a small absorbing target of arbitrary shape inside an
elongated domain of a slowly varying axisymmetric profile.  For this
purpose, the original Poisson equation in three dimensions is reduced
an effective one-dimensional problem on an interval with a
semi-permeable semi-absorbing membrane.  The approximate formula
captures correctly the dependence of the MFPT on the distance to the
target, the radial profile of the domain, and the size and the shape
of the target.  This approximation is validated by Monte Carlo
simulations.
\end{abstract}





\maketitle

\section{Introduction}

The concept of first-passage time, i.e., a time taken for a diffusing
particle to arrive at a given location, is very common in describing
many natural phenomena.  Nowadays it is widely used in chemistry
(geometry-controlled kinetics), biology (gene transcription, foraging
behavior of animals) and many applications (financial modelling,
forecasting of extreme events in the environment, time to failure of
complex devices and machinery, military operations), see
\cite{Redner_2001,Metzler_2014,Lindenberg_2019,Reguera_2001,Lanoiselee18,Grebenkov20a,Bressloff_2015,Reva_2020,Grebenkov_2016,Benichou_2014,Kalinay_2006,Rubi_2010,Lindsay_2017,Holcman_2015,Koopman_1980,Benichou_2011,Oshanin_2009,Benichou10,Benichou10b,Grebenkov18}
and references therein.

Most former works were dedicated to the {\it mean} first-passage time
(MFPT), which is also related the overall reaction rate on the target
region.  Since exact formulas for the MFPT are only available for a
few special cases of highly symmetric domains (such as sphere or
disk), a variety of powerful methods have been developed.  In
particular, many approximate solutions were derived in the so-called
narrow escape limit when the target size goes to $0$
\cite{Grigoriev02,Kolokolnikov05,Singer06a,Singer06b,Singer06c,Pillay10,Cheviakov10,Cheviakov11,Cheviakov12,Caginalp12,Holcman14,Marshall16,Grebenkov17b,Grebenkov18a,Grebenkov19}.  
While these asymptotic results are valid for generic domains, their
accuracy can be considerably reduced when the confining domain is
elongated (e.g., a long truncated cylinder or a prolate spheroid).  In
this case, the target region can still be very small as compared to
the diameter of the confining domain (i.e., the size of the domain
along the longitudinal direction), but comparable to the size of the
domain in the transverse directions.  The effect of the confinement
anisotropy onto the MPFT was studied in \cite{Grebenkov17d}.
Recently, we proposed a simple yet efficient method for deriving
approximate solutions of the MPFT in elongated domains on the plane
\cite{Grebenkov_2020}.  The aim of this paper is to extend this method
to three dimensions and to derive a general approximate formula for
the MFPT in an elongated three-dimensional domain with reflecting
boundaries.  The shape of the domain is assumed to be axisymmetric,
smooth and slowly varying in the longitudinal direction (without deep
pockets and enclaves), but otherwise general.  The target is assumed
to be small, but also of an arbitrary shape.  We validate our findings
by Monte Carlo simulations.

\section{Approximate solution} 

\begin{figure}
\begin{center}
\includegraphics[width=60mm]{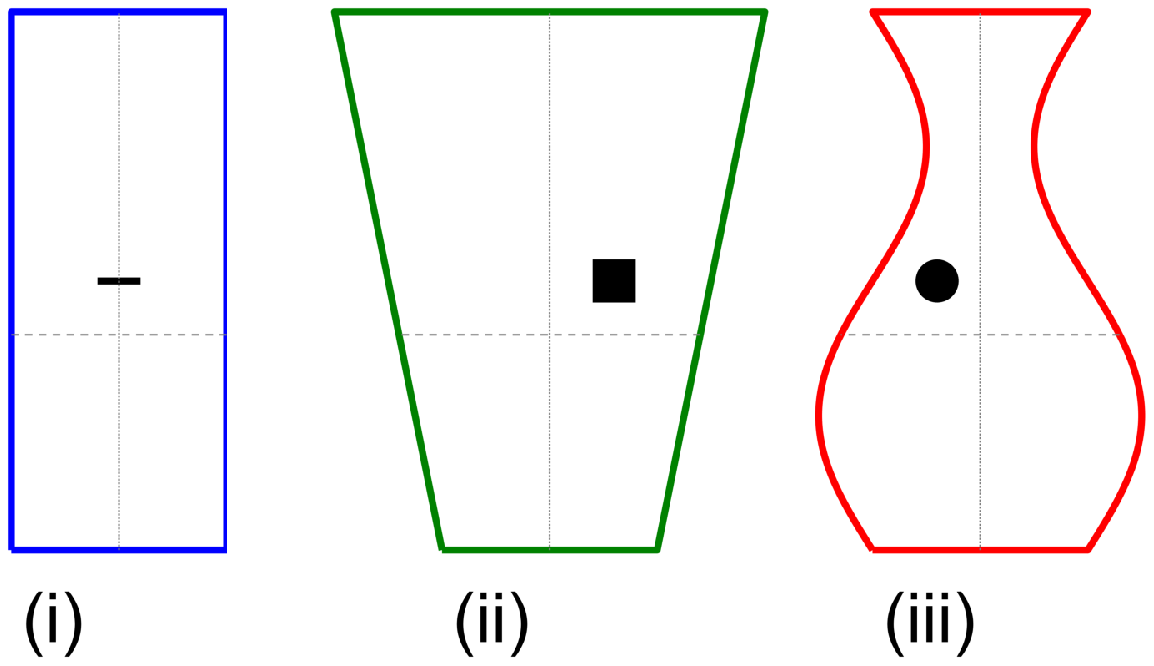} 
\hskip 5mm  \includegraphics[width=20mm]{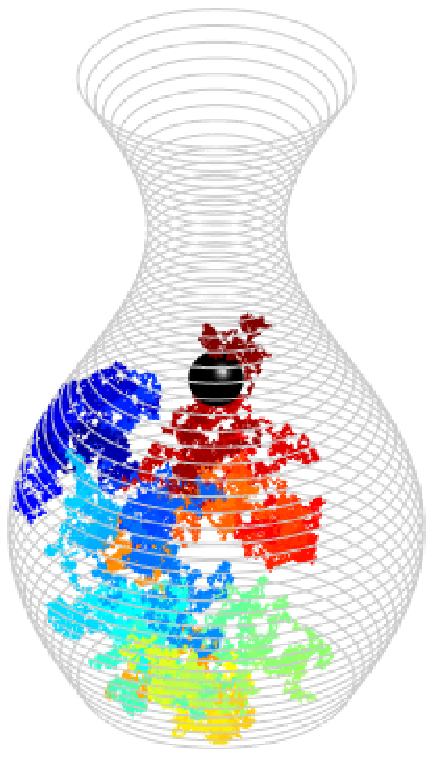} 
\end{center}
\caption{
Projection onto $xz$-plane of three domains used for Monte Carlo
simulations: truncated cylinder, $r(z) = 1$ (i), truncated cone, $r(z)
= 1 + z/\l$ (ii), and a domain with an oscillating profile $r(z) = 1 +
\tfrac12 \sin(2\pi z/\l)$ (iii), with $\l = 5$.  Vertical dotted line shows
the vertical symmetry axis along $z$ direction ($r=0$); horizontal
dashed line indicates the location of uniformly distributed starting
points (at $z = 2$).  A target (in black) is located at
$(x_T,0,\l/2)$: a disk of radius $\rho = 0.2$ with $x_T = 0$ (i), a
cube of edge $2\rho = 0.4$ with $x_T = 0.6$ (ii), and a sphere of
radius $\rho = 0.2$ with $x_T = -0.4$ (iii).  On the right, an example
of a simulated trajectory inside the domain with oscillating profile,
colored from dark blue to dark red according to elapsed time until the
first-passage to the target (black sphere) at the center.}
\label{fig:scheme}
\end{figure}

We consider an elongated axisymmetric domain of ``length'' $\l$, which
is determined by a smooth profile $r(z)$:
\begin{equation}
\Omega = \{(x,y,z)\in {\mathbb R}^3 ~:~ x^2 + y^2 < r^2(z), ~ 0<z<\l\}.
\end{equation}
Throughout the paper, we assume that the aspect ratio $r_0/\l$ of the
domain (with $r_0 = \max \{r(z)\}$) is small and its boundary profile
is smooth, $dr(z)/dz \ll 1$.  A small absorbing target is located
inside the domain at $(x_T,y_T,z_T)$, see Fig. \ref{fig:scheme}.

Similar to planar domains \cite{Grebenkov_2020}, the main analytical
formula will be derived by employing a three-step approximation.
First, the absorbing target is replaced by an absorbing disk of the
same trapping coefficient $K$; the disk is oriented perpendicular to
the symmetry axis of the domain.  Far away from the target such a
replacement is justifiable because at the distance greater than the
size of the target (but still much smaller than $r(z_T)$ and $\l$) the
absorption flux can be characterized by the first (monopole) moment of
the shape of the target, and this equivalence simply preserves it.
The trapping coefficient is proportional to the electrostatic
capacitance $C$ of the target, $K = 4 \pi D C$, where $D$ is the
diffusion coefficient \cite{Berezhkovskii_2007,Krapivsky_2010}.  For a
variety of shapes (e.g., sphere, ellipsoid, cube, prism, perturbed
axisymmetric shapes, or even some fractals objects), capacitance is
well-known or can be accurately estimated from various approximations,
see
\cite{Landau_1984,Chow_1982,Berezhkovskii_2007,Skvortsov_2018,Landkof_1978,Nair_2007,Wintle_2004,Piazza19}
and references therein.  For a disk of radius $a$, the capacitance is
$(2/\pi) a$ \cite{Landau_1984}.  Knowing the capacitance $C$ of a
given target shape, one can thus easily deduce the radius $a =
(\pi/2)C$ of the equivalent absorbing disk.

Second, we introduce the semi-permeable and semi-absorbing vertical
boundary (membrane) across the domain that passes through the
equivalent absorbing disk, i.e., at $z = z_T$, where $z_T$ is the
longitudinal target location.  In line with the conventional arguments
of effective medium theory, the trapping of the target can
approximately be captured by means of this boundary with some
effective reactivity $\kappa$.  A similar approach, often referred to
as the lump parameter approximation, has been applied in many areas of
physics and engineering (effective acoustic impedance of perforated
screens \cite{Crocker_1998}, effective electric conductance of
lattices and grids \cite{Tretyakov_2003}, effective boundary condition
for porous materials \cite{Hewett_2016,Marigo_2016,Martin_2022}).  To
relate the effective trapping rate of the membrane with the
geometrical setting, we assume that the effective trapping rate of the
membrane is equal to the trapping flux of the particles induced by the
presence of the target:
\begin{equation} \label{M:e20a} 
\kappa = \frac{K(r_T)}{S(z_T)} \,,
\end{equation}
where $S(z) = \pi r^2(z)$ is the cross-sectional area at ``height''
$z$.  We stress that $K$ and thus $\kappa$ depend on the radial
position $r_T = \sqrt{x_T^2 + y_T^2}$ of the target (an equivalent
disk) in the cross-section of the domain.  In other words, the
trapping coefficient $K(r_T)$ of the target inside the confining
domain is different from its value $K_0$ in the open space (when
$\Omega = {\mathbb R}^3$).  Moreover, it is the latter dependence that
determines the MFPT properties.  Calculation of the position-dependent
trapping coefficient $K$ is one of the main ingredients of the
proposed method.  In Appendix, we proposed the following approximation
\begin{equation} \label{M:e200} 
K = K_0\, \Psi \bigl(a/r(z_T), r_T/r(z_T)\bigr),
\end{equation}
where the function $\Psi (\nu, \eta)$, defined by Eq. (\ref{eq:Psi}),
was deduced by interpolating two analytical results for $r_T = 0$ (at
the symmetry axis of the domain), and for $r_T = R - a$ (near the
domain wall).  This function accounts for the relative target size
$\nu = a/r(z_T)$ and the relative traversal deviation $\eta =
r_T/r(z_T)$ of the target from the center of the domain cross-section.

Third, after its release at some point in the elongated domain, a
Brownian particle frequently bounces from the reflecting walls while
gradually diffusing along the domain towards the target.  The shape of
the walls (defined by $r(z)$) can additionally create the so-called
entropic drift, which can either speed up or slow down the arrival to
the target \cite{Reguera_2001,Kalinay_2006,Rubi_2010}.  In any case,
the information about the particle initial lateral location (e.g.,
across the domain) becomes rapidly irrelevant, and the original MFPT
problem, governed by the Poisson equation, is essentially reduced to
the one-dimensional problem.  While the classical Fick-Jacobs equation
determines the concentration of particles averaged over the
cross-section of the tube (see
\cite{Reguera_2001,Kalinay_2006,Rubi_2010} and references therein),
the survival probability is determined by the backward diffusion
equation with the adjoint diffusion operator \cite{Gardiner_1985}.  In
particular, the MFPT $T(z)$ in an elongated domain satisfies
\cite{Reguera_2001,Kalinay_2006,Rubi_2010,Lindsay_2017}
\begin{equation}  \label{FJ}
\frac{d}{dz} \biggl[S(z) \frac{dT}{dz} \biggr] = -\frac{S(z)}{D} \,.
\end{equation}
As the results of these approximations, the original problem of
finding the MFPT to a small target of arbitrary shape in a general
elongated domain is reduced to the one-dimensional problem, which can
be solved analytically.

We sketch only the main steps of the solution, while the details in a
similar case of planar domains can be found in \cite{Grebenkov_2020}.
We search for the solution of Eq. (\ref{FJ}) in the intervals
$(0,z_T)$ and $(z_T,\l)$.  Integrating this equation over $z$ and
imposing Neumann (reflecting) boundary conditions at $z = 0$ and $z =
l$, we get
\begin{equation}  \label{eq:Tx0}
T(z) = \left\{ \begin{array}{l l} \displaystyle C_- - \int\nolimits_0^z dz' \frac{V(z')}{D S(z')} & (0 < z < z_T), \\
\displaystyle C_+ -  \int\nolimits_z^{\l} dz' \frac{V(l) - V(z')}{D S(z')} & (z_T < z < \l), \\ \end{array} \right.
\end{equation}
where $V(z) = \int\nolimits_0^z dz' \, S(z')$ is the volume of
(sub)domain restricted between $0$ and $z$.  The integration constants
$C_\pm$ are determined by imposing the effective semi-permeable
semi-absorbing boundary condition at the target location $z = z_T$:
\begin{equation} \label{M:e8}
T(z_T - 0) = T(z_T + 0),   
\end{equation}
\begin{equation}  \label{eq:Robin2}
D \left [ \frac{d T}{dz}(z_T + 0) - \frac{d T}{dz}(z_T - 0) \right] = \kappa \,T(z_T) ,
\end{equation}
where $\kappa$ is given by  Eq. (\ref{M:e20a}). 
The first relation ensures the continuity of the MFPT, whereas the
second condition states that the difference of the diffusion fluxes at
two sides of the semi-permeable boundary at $z_T$ is equal to the
reaction flux on the target (an equivalent disk).  The latter flux is
proportional to $T$, with an effective reactivity $\kappa$ equal to
the effective trapping rate of the target, Eq. (\ref{M:e20a}).
Finally, substituting Eq. (\ref{eq:Tx0}) into Eqs. (\ref{M:e8},
\ref{eq:Robin2}), we get the solution of the problem:
\begin{equation}  \label{eq:Tx_main}
T(z) = \frac{l^2}{D} \biggl[{U}_{\sigma}(z_T/\l) - {U}_{\sigma}(z/\l)\biggr] + \frac{l}{\kappa} \, \frac{v(1)}{s(z_T/\l)} \, ,
\end{equation} 
where we introduced the following dimensionless quantities
\begin{equation} \label{eq:U}
{U}_{-}(\zeta) = \int\limits_0^\zeta d\zeta' \, \frac{v(\zeta')}{s(\zeta')} , \quad
{U}_{+}(\zeta) = \int\limits_\zeta^1 d\zeta' \, \frac{v(1) - v(\zeta')}{s(\zeta')} \,,
\end{equation}
with
\begin{eqnarray} 
\zeta &=& z/\l, \quad S(z) = \pi r_0^2 \, s(z/\l), \\   \label{eq:V}
V(z) &=& \pi r_0^2 \l \, v(z/\l), \quad v(\zeta) = \int\limits_0^{\zeta} d\zeta' \, s(\zeta').
\end{eqnarray}
The index $\sigma$ in Eq. (\ref{eq:Tx_main}) is the sign of $z - z_T$,
i.e., $\sigma = +$ for $z > z_T$, and $\sigma = -$ for $z < z_T$.
For a given profile $r(z)$, all these functions can be easily computed
either analytically (see examples in Table \ref{tab:Domains}), or
numerically.  In the simplest case of the cylindrical domain, $r(z) =
r_0$, one simply gets
\begin{equation} \label{M:e21}
T(z) = \left\{ \begin{array}{l l} \displaystyle \frac{z_T^2 - z^2}{2D} + \frac{\l}{\kappa } & ~~(0 \leq z \leq z_T), \\
 \displaystyle  \frac{(z - z_T)(2\l - z_T - z)}{2D} + \frac{\l}{\kappa } & ~~(z_T \leq z \leq \l). \\ \end{array} \right. 
\end{equation}

Equation (\ref{eq:Tx_main}) is the main result of the paper. As for
the case of planar domains \cite{Grebenkov_2020}, this equation
consists of two terms.  The first (diffusion) term is independent of
the size of the target and is related to the time required for a
Brownian particle to arrive at the proximity of the target from its
initial position.  For this reason, the contribution of this term is
small when $z \approx z_T$, i.e. when the particle initial position is
near the target.  The second (reaction) term in Eq.~(\ref{eq:Tx_main})
describes the particle absorption by the target when the particle
starts in its vicinity.  As it is inversely proportional to the target
size, this term dominates in the limit of very small targets.  We note
that the dependence on the lateral width of the domain comes only
through the parameter $\kappa$.

\begin{table}[ht!]
\begin{center}
\begin{tabular}{|c|c|c|c|c|c|c|}  \hline
Domain & $\rho(\zeta)$ & $v(\zeta)$ & $U_-(\zeta)$ & $U_+(\zeta)$ & $c_0$ & $c(\zeta)$ \\
\hline cylinder & $1$ & $\zeta$ & $\frac{1}{2} \zeta^2$ & $\frac12 (1-\zeta)^2$ &
$\frac13$ & $1-\zeta$ \\ cone & $\zeta$ & $\frac13 \zeta^3$ & $\frac{1}{6} \zeta^2$ &
$\frac{2-3\zeta+\zeta^3}{6\zeta}$ & $\frac{1}{15}$ & $\frac{1-\zeta}{3\zeta}$ \\
paraboloid & $\zeta^2$ & $\frac{1}{5} \zeta^5$ & $\frac{1}{10} \zeta^2$ & $\frac{3\zeta^5-5\zeta^3+2}{30\zeta^3}$ 
& $\frac{1}{35}$ & $\frac{1-\zeta^3}{15\zeta^3}$ \\  \hline
\end{tabular}
\caption{
Three examples of symmetric elongated domains defined by setting $r(z)
= r_0 \rho(z/\l)$, where $\rho(\zeta)$ is the rescaled radial profile,
$\zeta = z/\l$, $\l$ is the length of the domain, $r_0 =
\max\{r(z)\}$; $v(\zeta)$ is the rescaled volume in Eq.  (\ref{eq:V}),
functions ${U}_{\pm}(\zeta)$ are given in Eq.(\ref{eq:U}).  Constant
$c_0$ and function $c(\zeta)$ are given by Eq. (\ref{eq:C0C1}).  Other
examples can be deduced from similar expressions for the planar case
\cite{Grebenkov_2020} due to the identity $\rho^2(\zeta) = h(\zeta)$
between the rescaled profile $\rho(\zeta)$ of a three-dimensional
domain and the rescaled profile $h(\zeta)$ of the analogous
two-dimensional domain.  }
\label{tab:Domains} 
\end{center}
\end{table}

In many applications, the starting point is not fixed but uniformly
distributed inside the domain.  In this case, one often uses to the
volume-averaged MFPT
\begin{equation} \label{M:e30} 
\overline{T} = \frac{1}{V(l)} \int\limits^\l_0  \pi r^2 (z) T(z) dz.
\end{equation}
By substituting Eq. (\ref{eq:Tx_main}) into this expression we arrive
at
\begin{equation}
\label{M:e31} 
\overline {T} = \frac{\l^2}{D} \bigl(c_0 + c(z_T/\l) \bigr) + \frac{\l}{\kappa} \, \frac{v(1)}{s(z_T/\l)} \,,
\end{equation}
with
\begin{equation} \label{eq:C0C1}
c_0 =  \int\limits_0^1 d\zeta \frac{v^2(\zeta)}{v(1) s(\zeta)}  \,, \quad 
c(\zeta) = \int\limits_{\zeta}^1 d\zeta' \frac{v(1)}{s(\zeta')} \,.
\end{equation}
Note also that
\begin{equation}
U_+(\zeta) = c(\zeta) - (U_-(1) - U_-(\zeta)).
\end{equation}

\section{Discussion}
 
We use Monte Carlo simulations to check the accuracy of the analytical
predictions given by Eq. (\ref{eq:Tx_main}) in three geometrical
settings illustrated in Fig. \ref{fig:scheme}: (i) a disk of radius
$\rho$ in a truncated cylinder; (ii) a cube of edge $2\rho$ in a
truncated cone; and (iii) a sphere of radius $\rho$ in an oscillating
profile.  The capacitances of these targets are respectively
$(2/\pi)\rho$, $(4/3)\rho$ \cite{Wintle_2004,Berezhkovskii_2007}, and
$\rho$, from which the radius $a$ of an effective disk takes the
values $\rho$, $(2\pi/3)\rho$ and $(\pi/2)\rho$, respectively.  The
target is located at $(r_T,0,\l/2)$, where $\l = 5$ is the length of
the confining domains.  In each simulation run, a particle was
released from a random point uniformly distributed in the
cross-section at $z_0 = 2$.  It undertakes independent Gaussian jumps
with the standard deviation $\sigma = \sqrt{2D\delta}$ along each
coordinate, where $D = 1$ and $\delta = 10^{-6}$ is the time step.
The particle is reflected normally on the boundary of the confining
domain.  The simulation run is stopped when the particle crossed the
target.  The first-passage time is estimated as $n\delta$, where $n$
is the number of steps until stopping.  The MFPT is obtained by
averaging over $1000$ runs.

The approximate solution for a truncated cylinder is given in
Eq. (\ref{M:e21}), while the general expression (\ref{eq:Tx_main}) is
used for two other domains.  In the case of a truncated cone $r(z) = a
+ bz$, the functions $U_{\pm}(\zeta)$ can also be found explicitly:
\begin{equation*}
U_-(\zeta) = \frac{\zeta^2(3+\alpha \zeta)}{6(1+\alpha \zeta)}  \,, \quad
U_+(\zeta) = \frac{(1-\zeta)^2(3+2\alpha-\alpha \zeta)}{6(1+\alpha \zeta)} \,, 
\end{equation*}
with $\alpha = b\l/a$.  In turn, for an oscillating profile, it is
easier to calculate $U_{\pm}(\zeta)$ directly from their definition
(\ref{eq:U}) via numerical integration.

\begin{figure}
\begin{center}
\includegraphics[width=85mm]{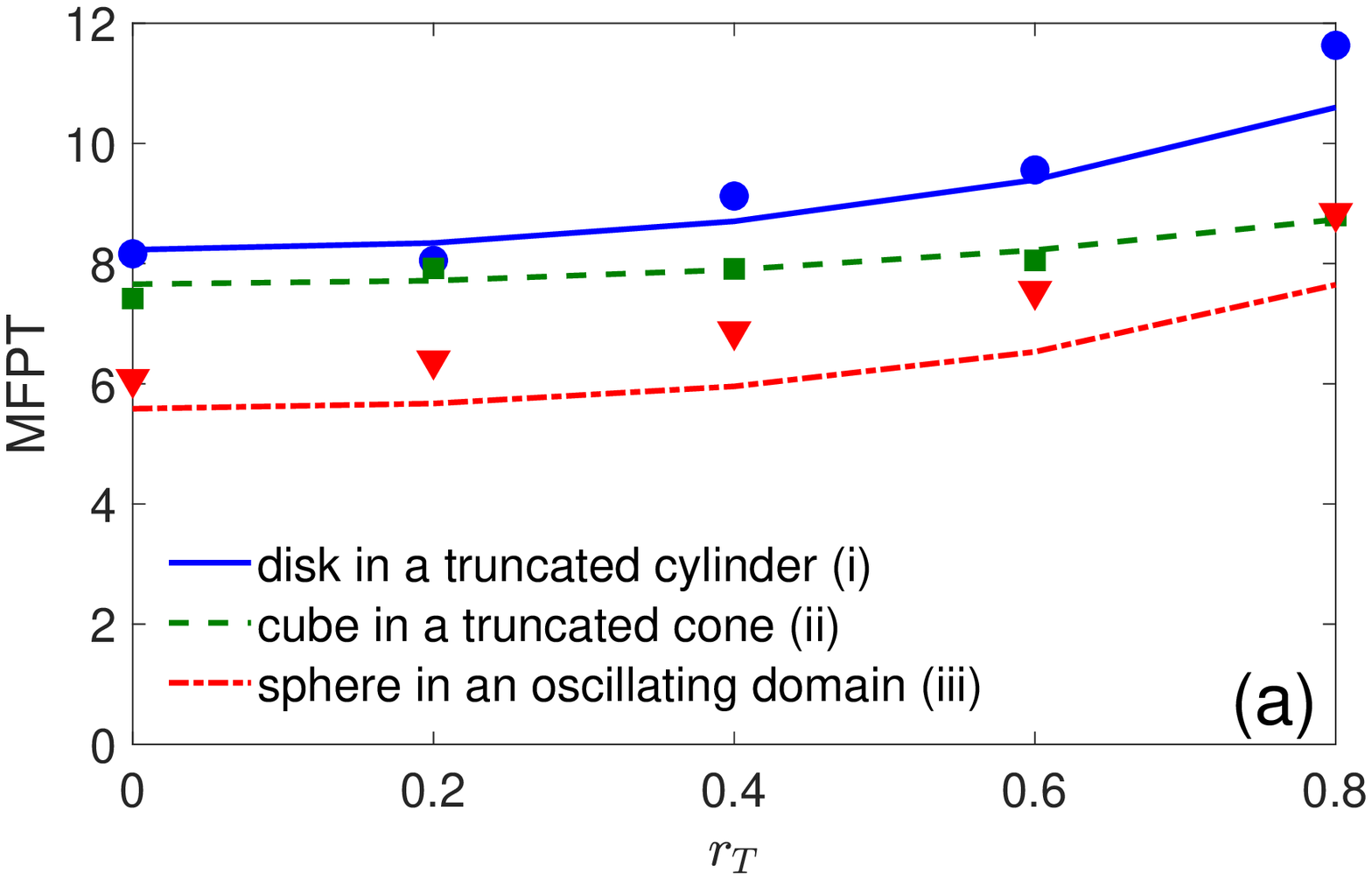} 
\includegraphics[width=85mm]{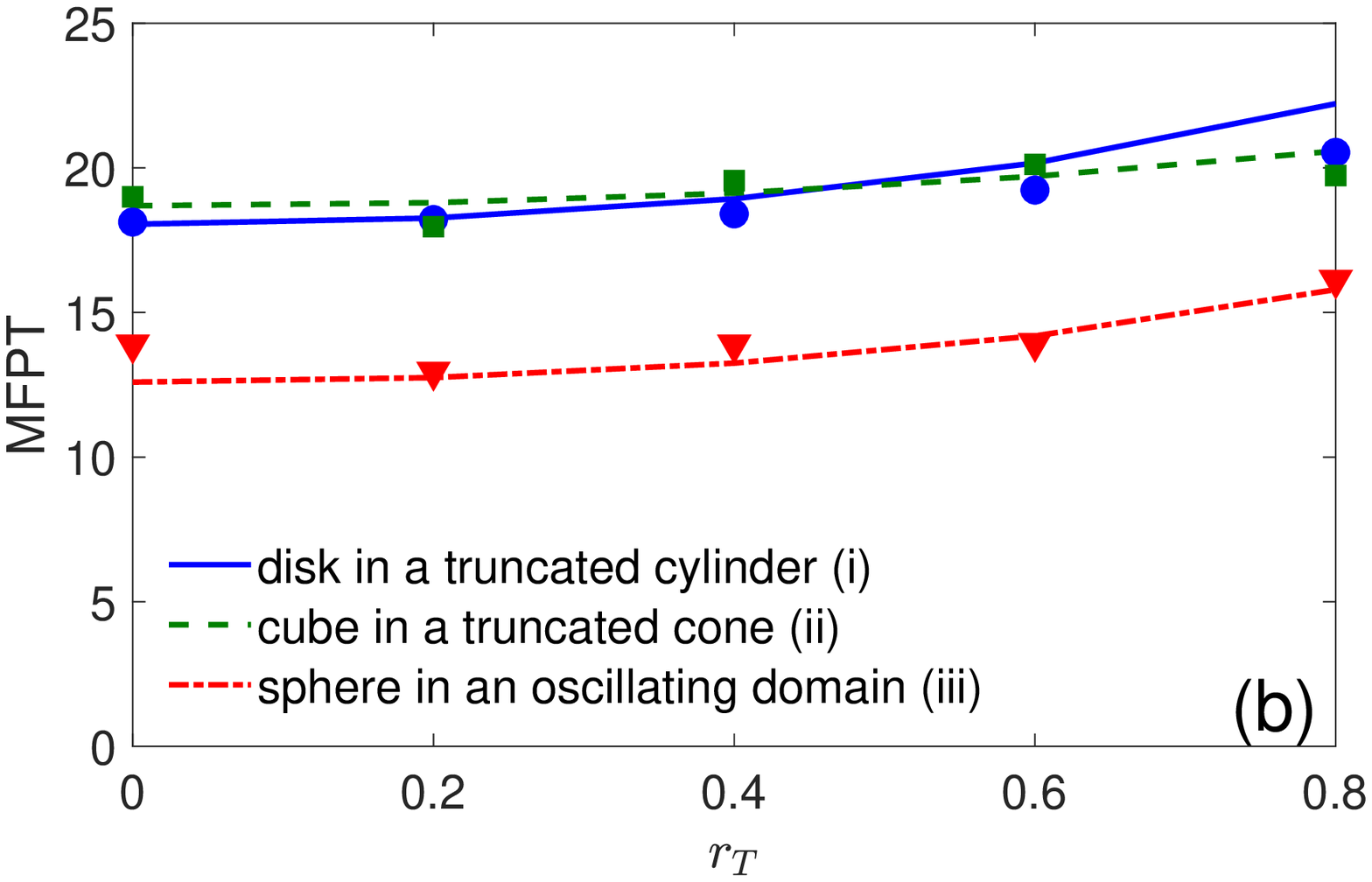} 
\end{center}
\caption{
MFPT as a function of $r_T$ for diffusion towards a target centered at
$\x_T=(r_T,0,\l/2)$, with $\rho = 0.2$ {\bf (a)} or $\rho = 0.1$ {\bf
(b)}, $D = 1$, $\l = 5$, the starting point $\x_0$ is uniform at the
cross-section at $z_0 = 2$, in three settings shown in
Fig. \ref{fig:scheme}: (i) a disk of radius $\rho$ inside a truncated
cylinder of radius $1$; (ii) a cube of edge $2\rho$ inside a truncated
cone $r(z) = 1 + z/\l$; and (iii) a sphere of radius $\rho$ inside an
oscillating domain $r(z) = 1 + \tfrac12 \sin(2\pi z/\l)$.  Lines show
theoretical predictions (\ref{eq:Tx_main}); symbols present the mean
values from $1000$ realizations obtained via Monte Carlo simulations
with the time step $\delta = 10^{-6}$.  }
\label{fig:MFPT}
\end{figure}

Figure \ref{fig:MFPT} presents the MFPT as a function of the radial
position $r_T$ of the target in the domain.  First of all, one can
note the overall agreement between our theoretical predictions and
Monte Carlo simulations.  Both theory and simulations indicate that
the MFPT increases when the target is shifted from the center ($r_T =
0$) towards the boundary of the confining domain ($r_T = 0.8$), even
though this effect is weak.  For a larger spherical target (panel (a),
$\rho = 0.2$), our approximation slightly underestimates the MFPT in
the case (iii) of an oscillating domain; the agreement is better for a
smaller target (panel (b), $\rho = 0.1$).  There are also minor
deviations for the case (i) when the disk is close to the boundary.
Despite these deviations, we conclude that our three-step
approximation accurately captures the properties of the MFPT in
elongated domains.  Given the simplistic character of this
approximation, its accuracy is striking.  It is worth stressing that
the targets are not too small (e.g., $\rho = 0.2$ is comparable to the
minimal radius of $r(0.75\l) = 0.5$ of the oscillating domain); the
domains are not too elongated (e.g., $r_0/\l = 0.4$ for the truncated
cone); and the particles are released not too far from the target
(here, $z_T - z_0 = 0.5$ is comparable to the target diameter $2\rho =
0.4$).  In other words, even though the assumptions of our
approximation are not fully satisfied, its predictions remain in a
quantitative agreement with Monte Carlo simulations.

\section{Conclusions and perspectives}

In this paper, we obtained a simple formula (\ref{eq:Tx_main}) for the
MFPT to a small absorbing target of an arbitrary shape in an elongated
axisymmetric domain with slowly changing boundary profile.  This
formula expresses the MFPT in terms of dimensions of the domain, the
form and te size of the absorbing target and its relative position
inside the domain.  We validated our analytical predictions by
numerical simulations and found excellent agreement.  Similar to the
planar domains \cite{Grebenkov_2020} the validity of the proposed
framework grounded on the condition of slowly changing profile $d
r(z)/d z \ll 1$.
 
A conventional way of improving the proposed approximation is to
account for the next order in the perturbation expansion, which
entails introduction of the position-dependent diffusion coefficient
\cite{Reguera_2001}
\begin{equation} \label{C2}
D \rightarrow \frac{D}{\sqrt{1 + [dr(z)/dz]^2}}.
\end{equation}
We note that under this approximation the results for the cylindrical
domain remain uncharged while an extension of the main formula
(\ref{eq:Tx_main}) is getting more challenging.

Future work may involve an extension of the proposed framework to more
complex geometries (an elongated domain with a compound piecewise
profile) or an extension to the slightly bended domain (but still with
a circular cross-section).  These extensions are straightforward; the
latter case reduces to a simple change of the coordinate $z$ in the
main equation (\ref{eq:Tx_main}) to the longitudinal curvilinear
coordinate along the bended domain.  The generalization of
Eq. (\ref{eq:Tx_main}) to domains with non-circular cross-section is
also possible, but is more involved and would require a substantial
refinement of relation (\ref{M:e200}), while the main
Eq. (\ref{eq:Tx_main}) remains valid.

We believe that the proposed expression for the MFPT is a useful tool
for some rapid practical estimations as well as for validation of
complex numerical models of particle diffusion in geometrically
constrained settings.

\begin{acknowledgments}
D.~S.~G. acknowledges a partial financial support from the Alexander
von Humboldt Foundation through a Bessel Research Award.
A.~T.~S. thanks Paul A. Martin for many helpful discussions.
\end{acknowledgments}

\appendix
\section{Effective trapping coefficient for an absorbing disk inside a tube}

We derive an approximate expression for the trapping coefficient $K$
of a small disk of radius $a$ in a reflecting tube with the
cross-sectional area $S(z) = \pi r^2(z)$.
For planar domains, the expression for $K$ can be deduced analytically
\cite{Grebenkov_2020}.  Unfortunately, there is no closed-form
analytical solution for $K$ in the case of a general position of the
absorbing disk in a three-dimensional tube with reflecting walls (we
note that the classical results for the capacitance of a small
conductor in a tube \cite{Smythe_1960,Smythe_1963,Chang_1970}
correspond to the Dirichlet boundary condition boundary condition on
the tube wall).  Nevertheless, there are some analytical results that
can be used to conjecture an accurate interpolating solution.

The expression for $K$ is indeed position dependent: $K = K(r_T,
z_T)$.  The dependence on $z_T$ is ``adiabatic'' and comes with the
slowly changing profile of the domain, $r(z)$.  As a function of
$r_T$, $K$ has a weak maximum at the center of the domain (the most
symmetrical configuration) that follows from the symmetry of the
problem and general bounds on the capacitance.  To capture these
properties we can begin with a simple ansatz
\begin{equation} \label{B:1}
\frac{K (\nu, \eta)}{K_0}   = A(\nu) [1 - B(\nu)  \eta^p],
\end{equation}
where $K_0 = 8aD$ is the trapping coefficient of a disk of radius $a$,
$\eta = r_T/r(z_T) \le 1$, $\nu = a/r(z_T) \ll 1$ and parameters
$A(\nu), B(\nu)$ and $p$ to be determined.

For $\eta =0$ (the centered disk) the solution has been derived by
Fock \cite{Fock_1941}, from which
\begin{equation}
\label{A:1a}
 A(\nu) =\frac{1 + 1.37 \nu - 0.37 \nu^4 }{(1 - \nu^2)^2} \ge 1. 
\end{equation}

The second parameter, $B(\nu)$, can be found from the situation when
the disk touches the wall of the tube.  In this case $\eta = 1 - \nu$
and we can write this condition in the form
\begin{equation} \label{A:1b}
\frac{K}{K_0 } = q A, 
\end{equation}
with some constant factor $q$. The value of factor $q$ can be deduced
from a general scaling argument.  It is well known that the
capacitance (and hence the trapping rate) scales with the square root
of the surface area of conductor (absorber)
\cite{Berezhkovskii_2007,Chow_1982}.  So the capacitance of any
conductor touching the reflecting wall is approximately $\sqrt{2}/2
\approx 0.71$ of its value at the center of the tube (at $\eta =0$),
which leads to Eq. (\ref{A:1b}).  This conjecture can also be
validated with the analytical results for two touching disks, $q =
3/4$ \cite{Fabrikant_1985} or $q = 0.74$ \cite{Strieder_2008}, and two
touching spheres when $q = \ln 2 \approx 0.69$
\cite{Berezhkovskii_2007,Lekner_2011}, which are reasonably close.
From here we arrive at
\begin{equation} \label{A:1c}
 B(\nu) = \frac{1 - q}{(1 - \nu)^p} > 0.
\end{equation}

The value of exponent $p =2$ can be deduced from comparison with the
analytical results for the flux of a monopole source in the tube
\cite{Martin_2022} near the tube center ($\eta =0$).  One can also
estimate it directly from numerical simulations.

Combining Eqs. (\ref{B:1}, \ref{A:1a}, \ref{A:1c}), we get our
analytical model (\ref{M:e200}) for the trapping coefficient, with
\begin{equation}  \label{eq:Psi}
\Psi (\nu, \eta) = A(\nu) [1 - B(\nu) \eta^2].
\end{equation}

\end{document}